\def\be{\begin{equation}}
\def\ee{\end{equation}}
\def\ba{\begin{array}}
\def\ea{\end{array}}

\documentclass[aps,amsmath,amssymb,amsfonts,pra]{revtex4}
\usepackage{graphicx}
\usepackage{epstopdf}
\def\qed{\leavevmode\unskip\penalty9999 \hbox{}\nobreak\hfill
     \quad\hbox{\leavevmode  \hbox to.77778em{%
               \hfil\vrule   \vbox to.675em%
               {\hrule width.6em\vfil\hrule}\vrule\hfil}}
     \par\vskip3pt}

\newtheorem{theorem}{Theorem}

\newtheorem{lemma}{Lemma}
\begin{document}

\title{Monogamy inequalities of entanglement of assistance in  $2\otimes 2\otimes d$ systems}

\author{Xue-Na Zhu$^{1}$}
\thanks{jing\_feng1986@126.com}
\author{Gui Bao$^{1}$}
\thanks{baoguigui@163.com}
\author{Zhi-Xiang Jin$^{2}$}
\thanks{ jzxjinzhixiang@126.com}
\author{Shao-Ming Fei$^{3}$}
\thanks{feishm@cnu.edu.cn}
\author{Tao Li$^{4}$}
\thanks{litao@btbu.edu.cn}
\affiliation{$^1$School of Mathematics and Statistics Science, Ludong University, Yantai 264025, China\\
$^2$School of Computer Science and Technology, Dongguan University of Technology, Dongguan, 523808, China\\
$^3$School of Mathematical Sciences, Capital Normal University, Beijing 100048, China\\
$^4$School of Mathematics and Statistics, Beijing Technology and Business University, Beijing 100048, China}

\begin{abstract}
The monogamy relations characterize the distribution of quantum correlations
among the multipartite quantum systems. We study the monogamy relations of the entanglement of assistance in $2\otimes 2\otimes d$ systems. We present explicitly the relations satisfied by the concurrence, the tangle and the concurrence of assistance, which can be used to derive rigorous monogamy relations. Detailed examples are given to illustrate our results.
\end{abstract}

\maketitle

\section{Introduction}
Quantum entanglement \cite{QE1,QE2,QE3,QE4} is an essential feature of quantum mechanics, which distinguishes the quantum from
the classical world and can be used to carry out tasks that can not be completed classically.
The quantification of quantum entanglement is a central problem in quantum information theory \cite{RP}. Many different
entanglement measures motivated by different ideas have been proposed. The concurrence is a well-known measure of entanglement
given by Hill and Wootters \cite{au} and Wootters \cite{pr}. For a bipartite pure state $|\psi\rangle_{AB}$ in systems $A$ and $B$, its concurrence is given by \cite{s7,s8,af},
\begin{equation*}\label{CON}
C(|\psi\rangle_{AB})=\sqrt{S_2(\rho_A)},
\end{equation*}
where $\rho_A$ is the reduced density matrix obtained by tracing over the subsystem $B$,
$\rho_{A}=Tr_{B}(|\psi\rangle_{AB}\langle\psi|)$ and the linear entropy $S_2(\rho)=2(1-Tr(\rho^2))$.
The concurrence is extended to mixed states
$\rho_{AB}=\sum_{i}p_{i}|\psi _{i}\rangle \langle \psi _{i}|$,
$0< p_{i}\leq 1$, $\sum_{i}p_{i}=1$, by the convex roof extension,
\begin{equation*}\label{CONC}
C(\rho_{AB})=\min_{\{p_i,|\psi_i\rangle\}} \sum_i p_i C(|\psi_i\rangle_{AB}),
\end{equation*}
where the minimum is taken over all possible pure state decompositions of $\rho_{AB}$.
For a tripartite state $|\psi\rangle_{ABC}$, the concurrence of assistance
(CoA) is defined by\cite{77,TFS}
\begin{equation}\label{Ca}
C_a(|\psi\rangle_{ABC})\equiv C_a(\rho_{AB})
=\max_{\{p_i,|\psi_i\rangle_{AB}\}}\sum_ip_iC(|\psi_i\rangle_{AB}),
\end{equation}
for all possible ensemble realizations of
$\rho_{AB}=Tr_{C}(|\psi\rangle_{ABC}\langle\psi|)=\sum_i p_i |\psi_i\rangle_{AB} \langle \psi_i|$.
When $\rho_{AB}=|\psi\rangle_{AB}\langle \psi|$ is a pure state, then one has
$C(|\psi\rangle_{AB})=C_{a}(\rho_{AB})$.
CoA is an entanglement monotone for $2\otimes 2\otimes d$ pure states \cite{72}.

Unlike classical correlations, quantum entanglement could be
monogamous \cite{Vc,m3}. The entanglement between two any
parties may limit the entanglement of these two parts with the rest parts.
Based on the explicit expression of concurrence for arbitrary two-qubit
states, Coffman, Kundu and Wootters \cite{RP} derived the famous genuine three-qubit
entanglement monotone, three tangle and conjectured the Coffman-Kundu-Wootters
(CKW) inequality satisfied by the concurrence.
Osborne and Verstraete \cite{prl96220503,024304} proved the CKW inequality for
arbitrary state $\rho_{ABC}$ in $2\otimes 2\otimes 2^{n-2}$,
\begin{equation}\label{cm1}
 C^2(\rho_{A|BC})\geq C^2(\rho_{AB})+C^2(\rho_{AC}),
\end{equation}
which can be extended to the case of $n$-qubit mixed states $\rho_{AB_1...B_{n-1}}$ \cite{prl96220503},
$C^2(\rho_{A|B_1...B_{n-1}})\geq \sum_{i=1}^{n-1}C^2(\rho_{AB_i})$, where
$\rho_{AB}=Tr_C(\rho_{ABC})$, $\rho_{AC}=Tr_B(\rho_{ABC})$
and $\rho_{AB_i}=Tr_{B_1...B_{i-1}B_{i+1}...B_{n-1}}(\rho_{A|B_1...B_{n-1}})$, $i=1,...,n-1$.

Dual to the CKW monogamy inequality, the concurrence of assistance
(CoA) satisfies \cite{72}
 \begin{equation}\label{ca1}
 C^2(|\psi\rangle_{A|BC})\leq C_a^2(\rho_{AB})+C_a^2(\rho_{AC})
\end{equation}
for pure states $|\psi\rangle_{ABC}$ in $2\otimes 2\otimes 2$,
which is generalized to the case of $n-$qubit pure states $|\psi\rangle_{A|B_1...B_{n-1}}$
\cite{012108},
\begin{equation}\label{ca2}
C^2(|\psi\rangle_{A|B_1...B_{n-1}})\leq \sum_{i=1}^{n-1}C_a^2(\rho_{AB_i}).
\end{equation}

Especially, it can be readily proved that the following
monogamy equality holds for a pure three-qubit state $|\psi\rangle_{ABC}$ \cite{Vc,81042306},
\begin{equation}\label{3de}
C^2(|\psi\rangle_{A|BC})= C^2(\rho_{AB})+C^2_a(\rho_{AC}).
\end{equation}
Instead of the inequality (\ref{cm1}), the equality
(\ref{3de}) gives a more exact characterization of entanglement
distribution. It would be interesting if the
equality (\ref{3de}) holds for any higher dimensional tripartite
quantum systems.

For tripartite pure states in
arbitrary dimensions, Ref.\cite{PRA71042331} has demonstrated that an inequality analogous to that presented in Ref.\cite{Vc} holds.
Based on the definition of monogamy without inequalities \cite{PRA99042305},
any measure of entanglement that on pure bipartite states is given by a strictly
concave function of the reduced density matrix is monogamous on pure tripartite states \cite{PRA99042305}.
Furthermore,
the authors in \cite{JMP} considered the
equality (\ref{3de}) in $2\otimes 2\otimes d$ quantum systems with $d\geq2$, and obtained that
i) if $C(|\psi\rangle_{A|BC})=C_a(\rho_{AC})$, then $C(\rho_{AB})=0$;
ii) $C(|\psi\rangle_{A|BC})=C(\rho_{AB})$ if only and if  $C_a(\rho_{AC})=0$;
iii) while there exists a state in a $2\otimes2\otimes d$
such that $C(\rho_{AB})=0$, but $C(|\psi\rangle_{A|BC})>C_a(\rho_{AC})$.
In \cite{QE4,ca} the authors presented an elegant monogamy equation
for $2\otimes 2\otimes d$-dimensional quantum
pure states, which reveals the relation among the bipartite concurrence,
CoA and genuine tripartite entanglement. However, in $2\otimes 2\otimes d$  quantum
systems, the monogamy inequality in terms of
$C(|\varphi\rangle_{A|BC})$, $C(\rho_{AB})$ and $C_a(\rho_{AC})$
is still unknown \cite{JMP}.

In this paper, concerning the open problem of monogamy properties between the concurrence and the CoA in $2\otimes 2\otimes d$ systems, we make
a comparative study based on another entanglement measure
closely related to the concurrence, the tangle $T$ \cite{PRA72022309},
\begin{equation}\label{T}
T(\rho_{AB})=\min\sum_ip_iC^2(|\psi_i\rangle_{AB})
=\min\sum_ip_iS_2(Tr_B(|\psi_i\rangle_{AB}\langle\psi_i|)),
\end{equation}
defined also as a minimization over pure state decompositions of $\rho_{AB}=\sum_i p_i |\psi_i\rangle_{AB} \langle \psi_i|$.

\section{ Monogamy relations for $2\otimes 2\otimes d$ systems}

Any two qubit state $\rho_{AB}$ may be written as \cite{prl96220503}
$
\rho_{AB}=\varLambda\otimes I_B(|V_{B^{\prime}B}\rangle\langle V_{B^{\prime}B}|),
$
where $|V_{B^{\prime}B}\rangle$ is the symmetric two qubit purification of the
reduced density operator $\rho_{B}$ on an auxiliary qubit system $B^{\prime}$, and $\varLambda$ is a qubit channel from $B^{\prime}$ to $A$.
The action of a qubit channel $\varLambda$ on a single-qubit state is given by
\begin{equation*}
 \varLambda(\rho)=\frac{I_2+(L\vec{r_B}+l)\sigma}{2},
\end{equation*}
where $\rho=\frac{I_2+\vec{r}_B\sigma}{2}$, $I_2$ is the $2\times 2$ identity, $\vec{r_B}$ is the Bloch vector, $\sigma$ is the vector of Pauli operators, $L$ is a $3\times3$ real matrix and ${l}$ is a three-dimensional vector. In terms of measurement-based conditional density operators, the linear entropy version of the
classical correlation of bipartite states $\rho_{AB}$ is defined by \cite{prl96220503,prl88017901},
\begin{equation*}
 I_2^{\leftarrow}(\rho_{AB})=\max_{\{P_i\}}[S_2(\rho_A)-\sum_ip_iS_2(\rho^i_A)],
\end{equation*}
where the maximum is taken over all positive operator-valued measure (POVM)
${P_i}$ performed on the subsystem $B$, satisfying $\sum_iP_i^{\dagger}P_i=I$ with the probability of the $i$th measurement outcome, $p_i=Tr[(I_A\otimes P_i)\rho_{AB}(I_A\otimes P_i^{\dagger})]$,
$\rho^i_A=Tr_B[(I_A\otimes P_i)\rho_{AB}(I_A\otimes P_i^{\dagger})]/p_i$ is the conditional
states of the system A associated with the measurement outcome $i$, $I_A$ and $I$ are the corresponding identity operators. It can be directly checked that \cite{zhuxuenadiscord}
\begin{equation}\label{I2}
I^{\leftarrow}_2(\rho_{AB})=\frac{2}{d}\lambda_{max}(L^TL)S_2(\rho_B),
\end{equation}
where $\lambda_{max}(K)$ is the maximal singular value of $K$.

For any pure states $|\varphi\rangle_{ABC}$, it was noted that \cite{PRA80034301}
\begin{equation}\label{IT}
S_2(\rho_A)=I_2^{\leftarrow}(\rho_{AB})+T(\rho_{AC}),
\end{equation}
where $\rho_A=Tr_{BC}(|\varphi\rangle_{ABC}\langle\varphi|)$,
$\rho_{AB}=Tr_C(|\varphi\rangle_{ABC}\langle\varphi|)$ and $\rho_{AC}=Tr_B(|\varphi\rangle_{ABC}\langle\varphi|)$. It is
implied that \cite{prl96220503},
\begin{equation}\label{IC}
I_2^{\leftarrow}(\rho_{AB})\geq T(\rho_{AB})= C^2(\rho_{AB})
\end{equation}
for two qubit states $\rho_{AB}$.

\begin{lemma}\label{TH1}
For all two qubit states $\rho_{AB}$, we have
$I_2^{\leftarrow}(\rho_{AB})\leq C_a^2(\rho_{AB})$.
\end{lemma}

{\sf [Proof]} Let $\rho_{AB}=\sum_{i=0}^{3}\mu_i|\varphi_i\rangle\langle\varphi_i|$ be the
spectral decomposition of $\rho_{AB}$, where $\mu_i$ and $|\varphi_i\rangle$ are respectively the eigenvalues and eigenvectors of $\rho_{AB}$, $i=0,1,2,3$. The purified state $|\psi\rangle_{ABC}$ of $\rho_{AB}$ is a $2\otimes 2\otimes 4$ pure state,
\begin{equation*}
|\psi\rangle_{ABC}=\sum_{i=0}^{3}\sqrt{\mu_i}|\varphi_i\rangle|i\rangle,
\end{equation*}
where $\{|i\rangle,\, i=0,1,2,3\}$ is the orthonormal basis of the system $C$.

Consider the Schmidt decomposition of a general pure state
$|\psi\rangle_{ABC}$ in $2\otimes 2\otimes 4$ systems \cite{PRA81024305},
\begin{equation*}
|\psi\rangle_{ABC}=\sqrt{r_0}|\phi_0\rangle_{AB}|0\rangle_C
+\sqrt{r_1}|\phi_1\rangle_{AB}|1\rangle_C+\sqrt{r_2}|\phi_2\rangle_{AB}|2\rangle_C
+\sqrt{r_3}|\phi_3\rangle_{AB}|3\rangle_C,
\end{equation*}
where $r_i\in[0,1]$, $i=0,1,2,3$ and $\sum_ir_i=1$,
$|\phi_0\rangle, |\phi_1\rangle, |\phi_2\rangle$ and $|\phi_3\rangle$
are orthonormal to each other. According
to the Schmidt numbers of $\{|\phi_0\rangle, |\phi_1\rangle, |\phi_2\rangle, |\phi_3\rangle\}$, the state $|\psi\rangle_{ABC}$ can be categorized into three classes \cite{PRA81024305}:

Case (1). There are no rank-2 states in $\{|\phi_0\rangle, |\phi_1\rangle, |\phi_2\rangle,|\phi_3\rangle\}$.

Under proper local bases, $|\psi\rangle_{ABC}$ can be expressed
as \cite{PRA81024305}
\begin{equation*}
|\psi\rangle_{ABC}=\sqrt{r_0}|00\rangle_{AB}|0\rangle_C
+\sqrt{r_1}|01\rangle_{AB}|1\rangle_C+\sqrt{r_2}|10\rangle_{AB}|2\rangle_C
+\sqrt{r_3}|11\rangle_{AB}|3\rangle_C.
\end{equation*}
In this case $\rho_{AB}=r_0|00\rangle\langle00|+r_1|01\rangle\langle01|+
r_0|10\rangle\langle10|+r_1|11\rangle\langle11|$ is a mixture of separable states.
Therefore, $C_a(\rho_{AB})=2\sqrt{r_0r_3}+2\sqrt{r_1r_2}$,
$S_2(\rho_B)=4(r_0+r_2)(r_1+r_3)$
and
\begin{eqnarray}\nonumber
L=\begin{pmatrix}
0&0&0&\\
0&0&0&\\
0&0&\frac{r_0r_3-r_1r_2}{
(r_0+r_2)(r_1+r_3)}&\\
  \end{pmatrix}.\quad
\end{eqnarray}
Since $(r_0+r_2)(r_1+r_3)=r_0r_3+r_1r_2+(r_0r_1+r_2r_3)\geq r_0r_3+r_1r_2+2\sqrt{r_0r_1r_2r_3}=(\sqrt{r_0r_3}+\sqrt{r_1r_2})^2$, we have
$$\frac{|r_0r_3-r_1r_2|}{
(r_0+r_2)(r_1+r_3)} \sqrt{S_2(\rho_B)}=\frac{2|r_0r_3-r_1r_2|}{\sqrt{(r_0+r_2)(r_1+r_3)}}\leq
2|\sqrt{r_0r_3}-\sqrt{r_1r_2}|.
$$
Obviously, $I_2^{\leftarrow}(\rho_{AB})\leq C_{a}^2(\rho_{AB})$, which is independent
on the choice the basis of $B$.

Case (2). There are two Schmidt rank-2 states.

With proper local bases, $|\psi\rangle_{ABC}$ can be expressed
as \cite{PRA81024305}
\begin{equation*}
|\psi\rangle_{ABC}=\sqrt{r_0}|00\rangle_{AB}|0\rangle_C
+\sqrt{r_1}|11\rangle_{AB}|1\rangle_C+\sqrt{r_2}(\sqrt{p}|01\rangle_{AB}
+\sqrt{1-p}|10\rangle_{AB})|2\rangle_C
+\sqrt{r_3}(\sqrt{1-p}|01\rangle_{AB}-\sqrt{p}|10\rangle_{AB})|3\rangle_C.
\end{equation*}
For convenience, we denote $k_1=r_0+r_2(1-p)+r_3p$ and $k_2=r_1+r_2p+r_3(1-p)$.
By direct computation we have $C_a(\rho_{AB})=2\sqrt{r_0r_1}+2\sqrt{p(1-p)(r_2-r_3)^2+r_2r_3}$,
$S_2(\rho^2_B)=4k_1k_2$ and
\begin{eqnarray}\nonumber
L=\begin{pmatrix}
\frac{(r_2-r_3)\sqrt{p(1-p)}}{\sqrt{k_1k_2}}&0&0&\\
0&-\frac{(r_2-r_3)\sqrt{p(1-p)}}{\sqrt{k_1k_2}}&0&\\
0&0&(\frac{r_0}{k_1}+\frac{r_1}{k_2}-1)&\\
  \end{pmatrix}.\quad
\end{eqnarray}
We verify that $|\frac{(r_2-r_3)\sqrt{p(1-p)}}{\sqrt{k_1k_2}}|\sqrt{S_2(\rho_B)}
=2|r_2-r_3|\sqrt{p(1-p)}\leq C_a(\rho_{AB})$ and
\begin{eqnarray*}
|\frac{r_0}{k_1}+\frac{r_1}{k_2}-1|\sqrt{S_2(\rho_B)}
&=&2\frac{|p(1-p)(r_2-r_3)^2+r_2r_3-r_0r_1|}{\sqrt{k_1k_2}}\\
&\leq&2|\sqrt{p(1-p)(r_2-r_3)^2+r_2r_3}-\sqrt{r_0r_1}|\\
&\leq&C_{a}(\rho_{AB}),
\end{eqnarray*}
where the first inequality is due to $k_1k_2=r_0r_1+r_0(r_2p+r_3(1-p))+r_1(r_2(1-p)+r_3p)+
(r_2p+r_3(1-p))(r_2(1-p)+r_3p)\geq r_0r_1+2\sqrt{r_0(r_2p+r_3(1-p))}
\sqrt{r_1(r_2(1-p)+r_3p)}+(r_2p+r_3(1-p))(r_2(1-p)+r_3p)
=(\sqrt{r_0r_1}+\sqrt{p(1-p)(r_2-r_3)^2+r_2r_3})^2.$
Thus we obtain $I_2^{\leftarrow}(\rho_{AB})\leq C_{a}^2(\rho_{AB})$.

Case (3). There are four Schmidt rank-2 states.

With proper local bases $|\psi\rangle_{ABC}$ can be expressed
as \cite{PRA81024305}
\begin{eqnarray*}
|\psi\rangle_{ABC}
&=&\sqrt{r_0}(\sqrt{p}|00\rangle_{AB}+\sqrt{1-p}|11\rangle_{AB})|0\rangle_C\\
&+&\sqrt{r_1}(\sqrt{1-p}|00\rangle_{AB}-\sqrt{p}|11\rangle_{AB})|1\rangle_C\\
&+&\sqrt{r_2}(\sqrt{q}|01\rangle_{AB}+\sqrt{1-q}e^{i\theta}|10\rangle_{AB})|2\rangle_C\\
&+&\sqrt{r_3}(\sqrt{1-q}|01\rangle_{AB}-\sqrt{q}e^{-i\theta}|10\rangle_{AB})|3\rangle_C.
\end{eqnarray*}
For convenience, we denote $t_1=r_0p+r_1(1-p)+r_2(1-q)+r_3q$, $t_2=r_0(1-p)+r_1p+r_2q+r_3(1-q)$, $L_{11}=\frac{(r_0-r_1)\sqrt{p(1-p)}+(r_2-r_3)\cos\theta\sqrt{q(1-q)}}{\sqrt{t_1t_2}},$
$L_{22}=\frac{(r_0-r_1)\sqrt{p(1-p)}-(r_2-r_3)\cos\theta\sqrt{q(1-q)}}{\sqrt{t_1t_2}},$
$L_{12}=L_{21}=\frac{(r_2+r_3)\sin\theta\sqrt{q(1-q)}}{\sqrt{t_1t_2}}$
and
\begin{equation*}
L_{33}=
\frac{r_0p+r_1(1-p)-r_2(1-q)-r_3q}{2t_1}+\frac{r_0(1-p)+r_1p-r_2q-r_3(1-q)}{2t_2}.
\end{equation*}
We have
$C_a(\rho_{AB})=2\sqrt{p(1-p)(r_0-r_1)^2+r_0r_1}+2\sqrt{q(1-q)(r_2-r_3)^2+r_2r_3},$
$S_2(\rho_B)=4t_1t_2$
and
 \begin{eqnarray}\nonumber
L=\begin{pmatrix}
L_{11}&L_{12}&0&\\
L_{21}&L_{22}&0&\\
0&0&L_{33}&\\
  \end{pmatrix}.\quad
\end{eqnarray}

Let $\lambda_i$ $(i=1, 2, 3)$ be the singular values of $L$.
We can show that
$$
|\lambda_{1,2}|=\left|\frac{L_{11}+L_{22}
\pm\sqrt{(L_{11}-L_{22})^2+4L_{12}L_{21}}}{2}\right|
\leq\frac{|L_{11}+L_{22}|+\sqrt{(L_{11}-L_{22})^2+4L_{12}L_{21}}}{2}.
$$
Then
\begin{eqnarray*}
|\lambda_{i}|\sqrt{S_2(\rho_B)}
&\leq&|2\sqrt{p(1-p)}(r_0-r_1)|+\sqrt{4q(1-q)((r_2-r_3)^2+4r_2r_3\sin^2\theta)}\\
&\leq&|2\sqrt{p(1-p)}(r_0-r_1)|+\sqrt{4q(1-q)(r_2-r_3)^2+4r_2r_3\sin^2\theta}\\
&\leq&|2\sqrt{p(1-p)}(r_0-r_1)|+2\sqrt{q(1-q)(r_2-r_3)^2+r_2r_3}\\
&\leq&C_{a}(\rho_{AB})
\end{eqnarray*}
for $i=1,2$, where the seconder inequality is due to $q(1-q)\leq (\frac{q+(1-q)}{2})^2=\frac{1}{4}.$
Moreover,
\begin{eqnarray*}
|\lambda_{3}|\sqrt{S_2(\rho_B)}
&=&
|L_{33}|\sqrt{S_2(\rho_B)}\\
&=&2\frac{|\left(p(1-p)(r_0-r_1)^2+r_0r_1\right)-\left(q(1-q)(r_2-r_3)^2+r_2r_3\right)|}{\sqrt{t_1t_2}}\\
&\leq&2|\sqrt{p(1-p)(r_0-r_1)^2+r_0r_1}-\sqrt{q(1-q)(r_2-r_3)^2+r_2r_3}|\\
&\leq&C_{a}(\rho_{AB}),
\end{eqnarray*}
where the first inequality is due to $t_1t_2\geq(\sqrt{p(1-p)(r_0-r_1)^2+r_0r_1}+\sqrt{q(1-q)(r_2-r_3)^2+r_2r_3})^2.$
Hence, we obtain $I_2^{\leftarrow}(\rho_{AB})\leq C_{a}^2(\rho_{AB})$.

With Cases (1), (2) and (3), we can conclude that, for a general state $\rho_{AB}\in 2\otimes 2$, its one-way correlation measure satisfies
$I_2^{\leftarrow}(\rho_{AB})\leq C_{a}^2(\rho_{AB}).$
$\Box$

From (\ref{IT}), (\ref{IC}) and Lemma \ref{TH1}, we obtain the following theorem.

\begin{theorem}\label{TH2}
For an arbitrary $2\otimes 2\otimes d$ state $|\psi\rangle_{ABC}$, the following monogamy inequalities hold,
\begin{equation}\label{cac}
C^2(\rho_{AB})+T(\rho_{AC})\leq C^2(|\psi\rangle_{A|BC})\leq C^2_a(\rho_{AB})+T(\rho_{AC}).
\end{equation}
\end{theorem}

Theorem \ref{TH2} gives a monogamy relation of entanglement
in terms of concurrence, CoA and tangle $ 2 \otimes 2 \otimes d$ tripartite pure states. This monogamy inequalities (\ref{cac}) show the monogamy of entanglement more explicitly than the CKW inequality and the monogamy inequality of $T_a$ defined by \cite{PRA80034301},
\begin{equation*}\label{Ta}
T_a(\rho_{AB})=\max\sum_ip_iC^2(|\phi_i\rangle_{AB})=\max\sum_ip_iS_2
(Tr_B(|\phi_i\rangle_{AB}\langle\phi_i|)),
\end{equation*}
where the maximum runs over all pure state decompositions of
$\rho_{AB}=\sum_i p_i |\phi_i\rangle_{AB} \langle \phi_i|$.
As the tangle of assistance satisfies $T_a(\rho)\geq \max\{C^2_a(\rho),T(\rho)\}$, we have $T_a(|\psi\rangle_{A|BC})=C^2(|\psi\rangle_{A|BC})\leq C_a^2(\rho_{AB})+T(\rho_{AC})\leq T_a(\rho_{AB})+T_a(\rho_{AC})$. From this
result, the monogamy inequality \cite{PRA80034301}
$T_a(\rho_{A|B_1...B_{n-1}})\leq\sum_{i=1}^{n-1}T_a(\rho_{AB_i})$ can be established for arbitrary $n$-qubit states $\rho_{AB_1...B_{n-1}}$.

For any $2\otimes 2\otimes d$ pure states $|\psi\rangle_{ABC}$, combining Theorem \ref{TH2} and the relationship between $C^2_a(\rho_{AC})$ and $T(\rho_{AC})$, we note that

(1) If $C^2_a(\rho_{AC})\geq T(\rho_{AC})$, then
\begin{equation}\label{22}
C^2(|\psi\rangle_{A|BC})\leq C^2_a(\rho_{AB})+ C^2_a(\rho_{AC});
\end{equation}

(2) If $C^2_a(\rho_{AC})\leq T(\rho_{AC}),$ then
\begin{equation}\label{33}
C^2(|\psi\rangle_{A|BC})\geq C^2(\rho_{AB})+ C^2_a(\rho_{AC}).
\end{equation}

For a quantitative analysis, we consider  the following examples to illustrate the our results.

\noindent{\it Example 1} Let us consider the $2\otimes 2\otimes 2$ pure state
$|\psi\rangle_{ABC}$. According to the definition of concurrence and CoA,
it is obvious that $C(\rho_{AC})\leq  C_a(\rho_{AC}).$
It has been proven in \cite{PRA72022309} that
$T(\rho)=C^2(\rho)$ for two qubit states $\rho$.
Analogous to Theorem \ref{TH2}, one can easily find that
$C^2(|\psi\rangle_{A|BC}) \leq C_a^2(\rho_{AB})+ C^2_a(\rho_{AC})$,
as introduced in \cite{72,012108}.

\noindent{\it Example 2}
We now take into account the following $2\otimes 2\otimes 3$
state \cite{JMP},
\begin{equation*}
|\varphi\rangle_{ABC}=\frac{1}{\sqrt{6}}|002\rangle+\frac{1}{\sqrt{3}}|100\rangle
+\frac{1}{\sqrt{6}}|112\rangle+\frac{1}{\sqrt{3}}|011\rangle.
\end{equation*}

It is clear that \cite{JMP} $C(|\varphi\rangle_{A|BC})=1$, $C(\rho_{AB})=0$ and
$C_a(\rho_{AC})=C(\rho_{AC})=\frac{2\sqrt{2}}{3}$.
According to the definition of concurrence and tangle,
one has $C^2(\rho_{AC})\leq T(\rho_{AC})$.
Analogous to  Theorem \ref{TH2}, we have $C^2(|\varphi\rangle_{A|BC})\geq C^2(\rho_{AB})+C^2_a(\rho_{AC}).$

Any given $n$-qubit quantum state can always be considered
as a $2\otimes 2\otimes d$ tripartite quantum state,
with $d$ denoting the total dimension of the $(n-2)$-qubit systems.
From Theorem \ref{TH2}, if $C^2_a(\rho_{AC})= T(\rho_{AC})$, the
inequalities (\ref{22}) and (\ref{33}) hold
simultaneously, see the following example.

\noindent{\it Example 3} Consider the following multi-qubit generalized W-class states \cite{WC,062306},
\begin{equation}\label{w}
|\varphi\rangle_{ABC_1...C_{n-2}}=a|000...0\rangle+b_1|10...0\rangle+...+b_n|00...1\rangle,
\end{equation}
where $|a|^2+\sum_{i=1}^{n}|b_i|^2=1$.
It is straightforward to derive \cite{WC} that
$C_a(\rho_{AB})=C(\rho_{AB})$ and $C_a(\rho_{AC_i})=C(\rho_{AC_i})(i=1,2,...,n-2)$.
Thereby, from Theorem \ref{TH2}, we have
\begin{equation}\label{W1}
C^2(|\varphi\rangle_{A|BC_1...C_{n-2}})=C^2(\rho_{AB})+T(\rho_{A|C_1...C_{n-2}}).
\end{equation}
From the CKW inequality (\ref{cm1}) and the inequality (\ref{ca1}) for $n$-qubit pure states, one has
\begin{equation}\label{W2}
C^2(|\varphi\rangle_{A|BC_1...C_{n-2}})=C^2(\rho_{AB})+\sum_{i=1}^{n-2}C^2(\rho_{AC_i}).
\end{equation}
Combining (\ref{W1}) and (\ref{W2}) we have
 \begin{equation*}\label{CT}
 T(\rho_{A|C_1...C_{n-2}})=\sum_{i=1}^{n-2}C^2(\rho_{AC_i}).
 \end{equation*}
The squared concurrence satisfies the inequality $\sum_{i=1}^{n-2}C^2(\rho_{AC_i}) \leq C^2(\rho_{A|C_1...C_{n-2}})\leq  T(\rho_{A|C_1...C_{n-2}})$, which implies that
\begin{equation}\label{TCC}
 T(\rho_{A|C_1...C_{n-2}})=C^2(\rho_{A|C_1...C_{n-2}}).
\end{equation}
Based on $C_a(\rho_{A|C_1...C_{n-2}})\geq C(\rho_{A|C_1...C_{n-2}})$ and $C(\rho_{AB})=C_a(\rho_{AB})$, it is obvious that $C^2(|\varphi\rangle_{A|BC_1...C_{n-2}})\leq C^2(\rho_{AB})+C^2_a(\rho_{A|C_1...C_{n-2}})$.

Furthermore, for any $(n-1)$-qubit subsystems $AC_{1}...C_{n-2}$ of $ABC_1...C_{n-2}$,
the reduced density matrix $\rho_{AC_{1}...C_{n-2}}$
of $|\varphi\rangle_{ABC_1...C_{n-2}}$ is a mixture of an $(n-1)$-qubit generalized
W-class state and the vacuum \cite{062306}.
By a straightforward calculation we obtain
$\rho_{AC_{1}...C_{n-2}}=|x\rangle_{AC_{1}...C_{n-2}}\langle x|
+|y\rangle_{AC_{1}...C_{n-2}}\langle y|,$
where
\begin{equation*}
|x\rangle_{AC_{1}...C_{n-2}}
=(a|00...0\rangle+b_1|10...0\rangle+b_{3}|01...0\rangle+b_{n}|00...1\rangle)
_{AC_{1}...C_{n-2}}
\end{equation*}
and
\begin{equation*}
|y\rangle_{AC_{1}...C_{n-2}}
=\sqrt{\sum_{k=3}^{n}\left|b_k\right|^2}|00...0\rangle
_{AC_{1}...C_{n-2}}
\end{equation*}
are the unnormalized $(n-1)$-qubit states in subsystems $AC_{1}...C_{n-2}$.
From the Hughston-Jozsa-Wootters theorem \cite{062306}, for any pure-state decomposition of
$\rho_{AC_{1}...C_{n-2}}
=\sum_{h=1}^{r}|\phi_h\rangle_{AC_{1}...C_{n-2}}\langle\phi_h|$, one has
$|\phi_h\rangle_{AC_{1}...C_{n-2}} =u_{h1}|x\rangle_{AC_{1}...C_{n-2}}+u_{h2}|y\rangle_{AC_{1}...C_{n-2}}$ for
some $r\times r$ unitary matrices $u_{h1}$ and $u_{h2}$ for each $h$.
The normalized state $|\tilde{\phi_h}\rangle_{AC_{1}...C_{n-2}}=|\phi_h\rangle_{AC_{1}...C_{n-2}}/\sqrt{p_h}$, $p_h=|\langle\phi_h|\phi_h\rangle|$, is a superposition of
an $n-1$qubit generalized W-class state and the vacuum, which
is again a generalized W-class state. Therefore, one has the concurrence of each $(n-1)$-qubit pure $|\tilde{\phi_h}\rangle_{AC_{1}...C_{n-2}}$,
\begin{equation}\nonumber
C^2(|\tilde{\phi_h}\rangle_{A|C_{1}...C_{n-2}})
=\frac{4}{p_h^2}|u_{h1}|^4|b_1|^2\sum_{k=3}^{n}|b_k|^2.
\end{equation}
Then for the $n-1$-qubit state $\rho_{AC_{1}...C_{n-2}}
=\sum_hp_h|\tilde{\phi_h}\rangle_{AC_{1}...C_{n-2}}\langle\tilde{\phi_h}|$, we have
\begin{equation}\nonumber
\sum_hp_hC(|\tilde{\phi_h}\rangle_{A|C_{1}...C_{n-2}})
=\sum_hp_h\frac{2}{p_h}|u_{h1}|^2|b_1|\sqrt{\sum_{k=3}^{n}|b_i|^2}
=2|b_1|\sqrt{\sum_{k=3}^{n}|b_i|^2}.
\end{equation}
Thus we obtain
\begin{eqnarray}\label{CCa}
C(\rho_{A|C_{1}...C_{n-2}})
&=&
\min_{\{p_h,|\tilde{\phi_h}\rangle_{A|C_{1}...C_{n-2}}\}}
\sum_hp_hC(|\tilde{\phi_h}\rangle_{A|C_{1}...C_{n-2}})\\[1mm]\nonumber
&=&\max_{\{p_h,|\tilde{\phi_h}\rangle_{A|C_{1}...C_{n-2}}\}}
\sum_hp_hC(|\tilde{\phi_h}\rangle_{AC_{1}...C_{n-2}})\\[1mm]\nonumber
&=&C_a(\rho_{A|C_{1}...C_{n-2}}).
\end{eqnarray}
Combining  Eq.(\ref{TCC}) and Eq.(\ref{CCa}), we obtain
 $C^2_a(\rho_{A|C_{1}...C_{n-2}})=T(\rho_{A|C_{1}...C_{n-2}})$.
From this formula and $C(\rho_{AB})=C_a(\rho_{AB})$, Theorem \ref{TH2} shows that $C^2(|\varphi\rangle_{A|BC_1...C_{n-2}})= C^2(\rho_{AB})+C^2_a(\rho_{A|C_1...C_{n-2}}).$

\section{Conclusions and Remarks}
We have presented two interesting monogamy inequalities satisfied by bipartite concurrence, CoA and tangle for $2\otimes 2\otimes d$ quantum pure states. These inequalities naturally lead to a genuine tripartite entanglement measure for $2\otimes 2\otimes d$ tripartite quantum pure
states from a new perspective. Up to now, the quantitative relation among $C(|\psi\rangle_{A|BC})$, $C(\rho_{AB})$ and
$C_a(\rho_{AC})$ in general tripartite systems has been an open problem.
For $2\otimes 2\otimes d$ quantum pure states $|\psi\rangle_{ABC}$, our Theorem \ref{TH2} provides analytical expressions, $C^2(|\psi\rangle_{A|BC})\geq C^2(\rho_{AB})+C^2_a(\rho_{AC})$   if $C^2_a(\rho_{AC})\leq T(\rho_{AB})$, and $C^2(|\psi\rangle_{A|BC})\leq C_a^2(\rho_{AB})+C^2_a(\rho_{AC})$  if $C^2_a(\rho_{AC})\geq T(\rho_{AB})$,
which reveal the relation among bipartite concurrence and its assistance.
Our results may shed new light on not only the monogamy of entanglement but also the
quantification of multipartite entanglement.

\bigskip
\noindent{\bf Acknowledgments}\, \,
This work is supported by the National Natural Science Foundation of China under
grant No. 12301582;  the research
award fund for Natural Science Foundation of Shandong Province under Grant No. ZR2024MA068; the specific research fund of the Innovation Platform for Academicians of Hainan Province under Grant No. YSPTZX202215;
Guangdong Basic and Applied Basic Research Foundation under Grants No. 2024A1515030023; the Start-up Funding of Dongguan University of Technology No. 221110084;  GDSTA: SKXRC2025442.

\end{document}